\documentclass[a4paper,11pt]{article}
\pdfoutput=1 

\usepackage{jcappub}
\usepackage[T1]{fontenc}
\usepackage{amsmath}
\usepackage{amssymb}
\usepackage{graphicx}
\usepackage{hyperref}
\usepackage{xcolor}
\usepackage{booktabs}
\usepackage[toc,page]{appendix}
\usepackage{etoolbox}
\usepackage[capitalize]{cleveref}
\usepackage{siunitx}

\crefname{appendix}{App.}{Apps.}
\Crefname{appendix}{Appendix}{Appendices}

\makeatletter
  \let\orig@appendix\appendix
  \renewcommand{\appendix}{%
      \orig@appendix
      \crefalias{section}{appendix}%
  }
\makeatother

\newcommand{\spore}{\textsc{spore}}

\title{\textsc{spore}: An Event-Level Sampling Pipeline for Multi-Telescope Neutrino Astronomy}

\author[a]{Jeffrey Lazar}
\author[a]{Perrine Wilmet}
\author[a]{Gwenha\"{e}l de Wasseige}

\emailAdd{jlazar@icecube.wisc.edu}
\emailAdd{gwenhael.dewasseige@uclouvain.be}

\affiliation[a]{Centre for Cosmology, Particle Physics and Phenomenology -- CP3, Universit\'{e} catholique de Louvain, Louvain-la-Neuve, Belgium}

\abstract{We present an open-source Python package for simulating neutrino events from astrophysical point and extended sources using tabulated instrument response functions (IRFs).
The package encodes three detector response components---effective area, point spread function, and energy resolution---in a selection-agnostic HDF5 format, which can represent a neutrino telescope whose response is supplied in that form, whether from a public release or a private study.
Sampling algorithms cover point sources (inverse-CDF with Poisson or fixed-count modes), extended sky distributions (hierarchical inverse-CDF sampling, including full RA- and declination-dependent flux maps), and multi-detector joint analyses.
We validate the framework via a round-trip consistency test using the publicly available IceCube 10-year tracks data release: the released IRFs are ingested into the package and used to generate a synthetic event set, whose declination distribution reproduces the observed one to 10--15\% across the northern sky.
The reconstructed-energy distribution agrees to within about a third over the bulk of the sample but exceeds the data by up to a factor of three below \SI{600}{\GeV}, a discrepancy we trace to the coarse true-energy binning of the public smearing matrix rather than to the sampling: an independent forward fold of the same IRFs reproduces it.
We further compare against the IceCube HESE 7.5-year public data release: the sampled deposited-energy spectrum tracks the published best-fit expectation, and the observed data fall within the goodness-of-fit distribution built from 1{,}000 sampled pseudo-experiments, though on its well-fitting side, as expected for an expectation that was itself fit to those data.}


\begin{document}
\maketitle

\section{Introduction}
\label{sec:intro}

High-energy astrophysical neutrinos are unique probes of the non-thermal Universe.
Because they travel undeflected from their production sites and interact only weakly, they carry directional and spectral information from environments that are opaque to photons: the cores of active galactic nuclei, the interiors of gamma-ray burst jets, compact binary merger remnants, and the dense molecular clouds that enshroud star-forming regions~\citep{Meszaros:2019xej}.
IceCube's discovery of a diffuse high-energy astrophysical neutrino flux~\citep{IceCube:2013low}, its subsequent characterization~\citep{IceCube:2020wum,IceCube:2025tgp}, and the first evidence for individual neutrino sources~\citep{IceCube:2018cha,IceCube:2022der} have established high-energy neutrino astronomy as a mature field.

The field is now entering two overlapping transition eras that raise the stakes for simulation tools.
The first is the \emph{multi-telescope era}: IceCube, KM3NeT~\citep{KM3Net:2016zxf}, P-ONE~\citep{P-ONE:2020ljt}, TRIDENT~\citep{TRIDENT:2022hql}, and IceCube-Gen2~\citep{IceCube-Gen2:2020qha} will operate simultaneously, each with distinct geographic locations, detection media, and angular and energy resolutions.
The second is the \emph{multi-energy era}: proposed detectors---TAMBO~\citep{Thompson:2023pnl,Arguelles:2026btb}, TRINITY~\citep{Otte:2019knb}, GRAND~\citep{GRAND:2018iaj}, and HERON~\citep{GRAND:2025rps}---will extend the observable neutrino spectrum by several decades in energy beyond current water- and ice-Cherenkov arrays.
Understanding how best to use the combined data across different fields of view and energy bands requires open-source tools that give access to event-level information across detectors.

For Cherenkov neutrino telescopes, the existing simulation landscape does not fill this role.
\textsc{Prometheus}~\citep{Lazar:2023gem} provides an end-to-end Monte Carlo pipeline, but processing the raw photon-hit output to final analysis level is computationally costly and requires access to proprietary reconstruction chains.
At the other end of the spectrum, \textsc{toise}~\citep{vanSanten:2022wss} accepts tabulated instrument response functions (IRFs) and returns expected counts or sensitivity curves---well suited to design studies, but operating on aggregate distributions rather than individual events.

In this work, we present \spore{}---Sampling Pipeline for Observatory Response Estimation\footnote{\url{https://github.com/jlazar17/spore}}---a Python-based package that closes this gap.
\spore{} reads tabulated IRFs in a standardized HDF5 format and generates Monte Carlo samples of individual reconstructed events with realistic PSF smearing, energy resolution, and morphology labeling.
Two source classes are supported natively---point sources (\cref{sec:algorithms:ps}) and spatially extended emission with arbitrary RA and declination structure (\cref{sec:algorithms:extended})---and both samplers accept heterogeneous detector lists directly, so simultaneous multi-detector joint analyses (\cref{sec:algorithms:multidet}) need no separate class.

In order to fill this gap, \spore{} satisfies three main criteria.
First, \spore{} is IRF-agnostic and operates directly from published IRFs; no proprietary simulation or reconstruction chain is required.
Second, the sampler returns a list of simulated events, each carrying true and reconstructed direction, energy, time, morphology, and local zenith and azimuth angles computed from the event time and detector location.
Finally, multiple detectors can be passed to either sampler at once; each independently Poisson-samples its event count, and the combined event list is tagged by detector, enabling joint analyses out of the box.
In this way, \spore{} simulates astrophysical and atmospheric neutrinos; while it is possible to model atmospheric muons in principle, no IRF is currently provided for this purpose.

The remainder of this paper is structured as follows.
\cref{sec:software} describes the software design.
\cref{sec:irf} describes the IRF framework.
\cref{sec:sources} presents the source models.
\cref{sec:algorithms} details the sampling algorithms.
\cref{sec:validation} presents the validation.
\cref{sec:conclusions} summarizes and outlines future directions.

\section{Software Design}
\label{sec:software}

\spore{} is a pure-Python package (requiring Python $\geq 3.11$) distributed under the LGPL-3.0-or-later license and installable via \texttt{pip} or \texttt{poetry}.
It is organized into five sub-packages: \texttt{conventions}, which handles coordinates and unit primitives; \texttt{physics}, which lays out the neutrino species and event morphology enumeration; \texttt{source}, which contains flux models and spatial descriptions; \texttt{detector}, which defines IRF loading and interpolation; and \texttt{event\_sampling}, which implements sampler classes and event representation.
A detailed description of each sub-package, the interpolation methods used for IRFs, and the TOML-based configuration system is given in \cref{app:architecture}.
The IRF storage format is described in \cref{app:hdf5}.

\begin{figure*}[t]
  \centering
  \includegraphics[width=\textwidth]{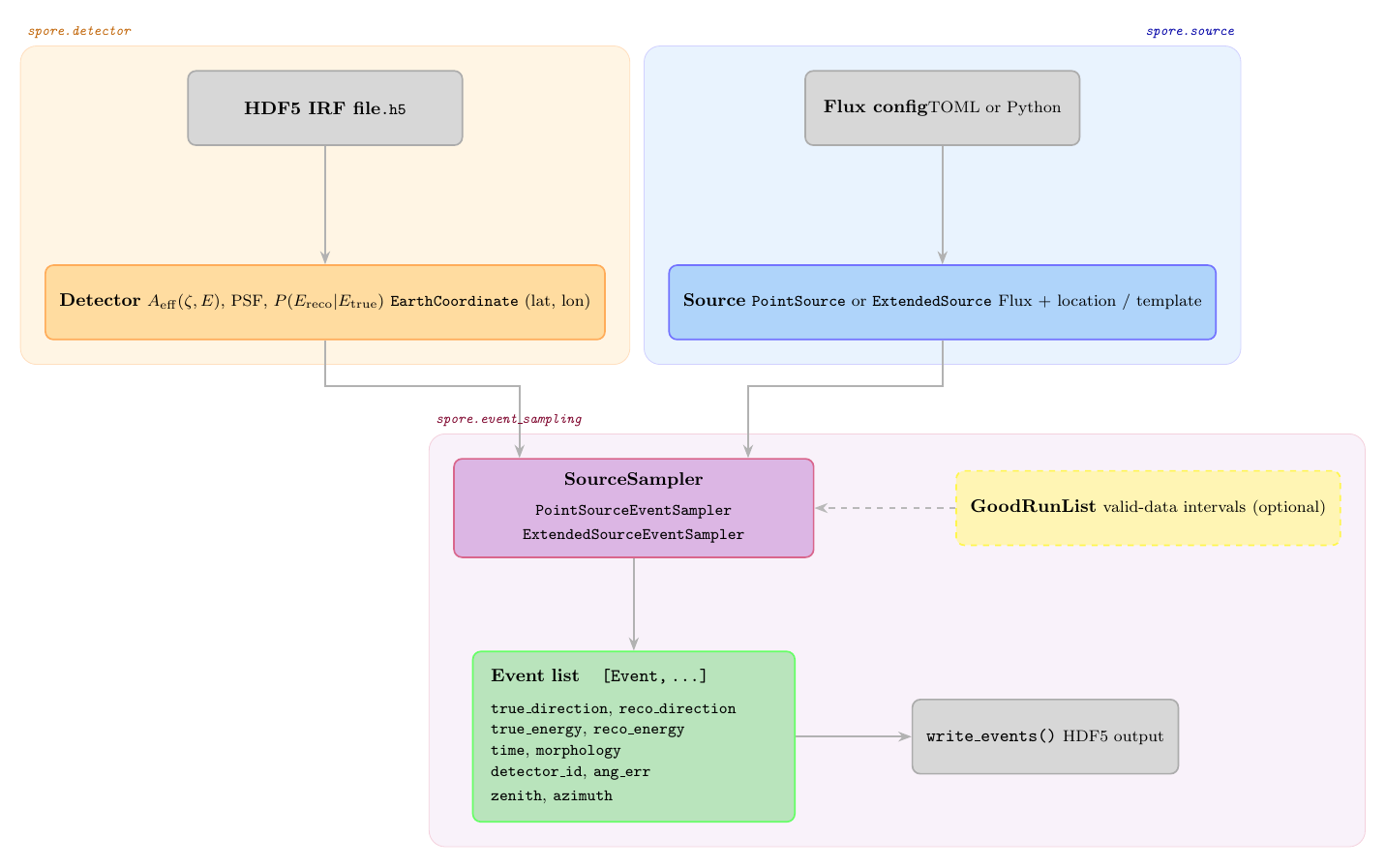}
  \caption{
    Package architecture.
    Arrows indicate data flow.
    A \texttt{Source} object specifies the astrophysical signal (flux and spatial distribution); a \texttt{Detector} object encapsulates the three IRF components loaded from an HDF5 file; a \texttt{Sampler} combines the two to produce a list of \texttt{Event} objects carrying true and reconstructed direction, energy, time, morphology label, detector identifier, and local zenith and azimuth angles.
    All components can be constructed programmatically or via TOML configuration files.
  }
  \label{fig:architecture}
\end{figure*}

The user provides two inputs: a \textbf{detector} and a \textbf{source}.
A detector is built from an HDF5 response file that contains the effective area, point-spread function, and energy resolution for each supported event morphology, as well as the detector's geographic latitude and longitude.
A source is built from a spectral flux model and a spatial distribution, i.e., a location for point sources or a spatial template for extended sources.
The flux model can be specified as a power-law (via a TOML configuration file or directly in Python), or as a tabulated spectrum loaded from an HDF5 file whose columns give energy in GeV and flux per neutrino species in \si{\GeV^{-1}\cm^{-2}\s^{-1}sr^{-1}}.
Two- and three-dimensional tabulated fluxes---functions of energy and declination, or of energy, declination, and right ascension---are also supported for extended sources and are loaded from HDF5 files with the corresponding additional axes.

These two inputs are combined in a sampler, which convolves the source flux with the detector response to produce a list of simulated events; \cref{fig:architecture} summarizes how the pieces fit together.
Each event carries a true direction and energy; a reconstructed direction and energy obtained by applying the PSF and energy resolution; an event time drawn within the observation livetime; a morphology label; and the local zenith and azimuth angles at the time and location of the observing detector, computed analytically from the event time via the local sidereal time.
When multiple detectors are provided, events from each are generated independently and tagged with an integer detector index before being returned in a single combined list.
The observation livetime, reference epoch, and energy smearing parameters can each be supplied as a single value shared across all detectors or as a per-detector list.
A good run list encoding discrete valid-data intervals may be supplied in place of a scalar livetime; event times are then drawn within those intervals, properly accounting for detector downtime.

\section{Instrument Response Function Framework}
\label{sec:irf}

In \spore{}, a neutrino detector's response to a neutrino source is described by three functions: the effective area $A_\mathrm{eff}(E, \delta)$, the point spread function $\mathrm{PSF}(\psi | E)$, and the energy resolution $P(E_\mathrm{reco} | E_\mathrm{true})$.
All three are stored as tabulated functions in a standard HDF5 format.
Where an IRF instead provides the joint distribution of the reconstructed quantities, that form is used directly in place of the two separable responses.
By default, \spore{} ships with IRFs for the IceCube ten-year point-source sample~\citep{IceCube:2021xar} and the HESE 7.5-year release~\citep{IceCube:2020wum}.

\subsection{Effective Area}
\label{sec:irf:effa}

The effective area, $A_\mathrm{eff}(\zeta,\, E_{\nu})$, quantifies the detector's sensitivity to a neutrino of energy $E_{\nu}$ arriving at zenith angle $\zeta$.
It is defined such that the expected number of detected events per unit time from a differential flux $\Phi \equiv \mathrm{d}N/(\mathrm{d}E\,\mathrm{d}A\,\mathrm{d}\Omega\,\mathrm{d}t)$ is
\begin{equation}
  \frac{\mathrm{d}N}{\mathrm{d}t} = \int A_\mathrm{eff}(\zeta(\delta,\alpha,t), E_{\nu})\,\Phi(\delta, \alpha, E_{\nu})\,\mathrm{d}E_{\nu}\,\mathrm{d}\Omega,
  \label{eq:rate}
\end{equation}
where the zenith angle follows from the source direction, the observation epoch, and the detector's geographic latitude $\phi$ and east longitude $\lambda$ as
\begin{equation}
  \cos\zeta = \sin\phi\,\sin\delta + \cos\phi\,\cos\delta\,\cos H,
  \qquad
  H = \mathrm{LST}(t, \lambda) - \alpha,
  \label{eq:zenith}
\end{equation}
with $H$ the hour angle of the source and $\mathrm{LST}$ the local sidereal time, which advances by one turn per sidereal day and is offset by the longitude.
The latitude therefore fixes the range of zenith angles a source can occupy, while the epoch and longitude together fix where in that range it sits at a given moment.
The package maintains separate effective areas for track-like events---predominantly from $\nu_\mu$ and $\bar\nu_\mu$ charged-current interactions that produce a long muon track---and cascade-like events---from all other interactions that produce a roughly spherical light deposition.
This distinction is important because the angular and energy resolutions differ by more than an order of magnitude between the two morphologies.

\begin{figure}[t]
  \centering
  \includegraphics[width=\textwidth]{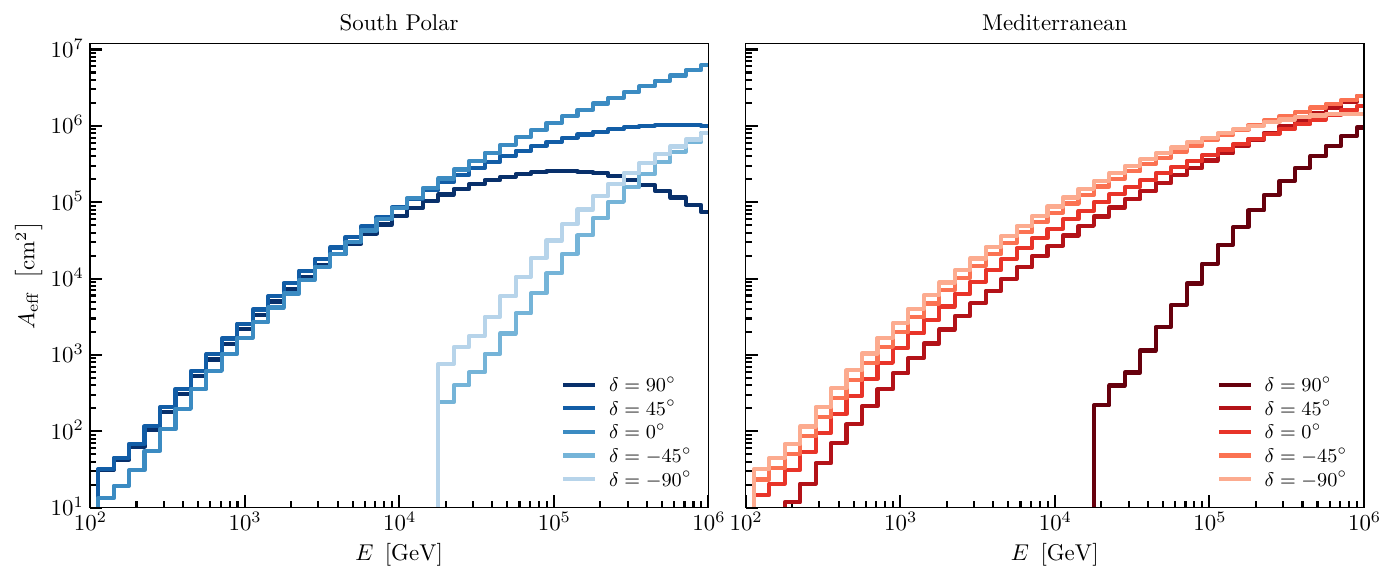}
  \caption{\textbf{\textit{Track-selection effective area for South Polar and Mediterranean detectors.}}
  Effective areas have been averaged over a day for both detectors.
  The trivial transform between local and equatorial coordinates for the South Polar detector results in more extreme variation between declinations.
  In contrast, the non-trivial transform results in more consistent effective areas for the Mediterranean detector (red).
  }
  \label{fig:effa_energy}
\end{figure}

\cref{fig:effa_energy} shows $A_\mathrm{eff}(E_{\nu})$ for track events, averaged over a full diurnal cycle, at five declinations, $\delta \in \{90^{\circ}, 45^{\circ}, 0^{\circ}, -45^{\circ}, -90^{\circ}\}$, for a South Polar detector and a Mediterranean detector at latitude $36.3^\circ$N.
In both cases, the effective area has been taken from the IceCube 10-year instrument response~\citep{IceCube:2021xar}.
The strong energy dependence reflects the interplay between the neutrino-nucleon cross section, the muon range, and the geometric acceptance.
For the South Polar detector, the mapping is especially simple: with $\phi = -90^{\circ}$ the hour-angle term in \cref{eq:zenith} vanishes, leaving $\cos\zeta = -\sin\delta$.
Each declination therefore corresponds to a single, time-independent zenith angle---overhead at $\delta = -90^{\circ}$, the horizon at $\delta = 0$, and straight through the Earth at $\delta = +90^{\circ}$, where absorption turns the response over above $\sim\SI{100}{\TeV}$.
The strong zenith dependence of the effective area thus maps directly onto declination, with no diurnal averaging to soften it.
The Mediterranean detector shows weaker declination variation because the non-trivial transformation between equatorial and local coordinates smears the zenith-angle distribution when averaging over the diurnal cycle.

\subsection{Point Spread Function}
\label{sec:irf:psf}

The point spread function (PSF) describes the conditional distribution of the angular offset $\psi$ between the true and reconstructed neutrino arrival direction:
\begin{equation}
  \mathrm{PSF}(\psi \mid E) = \frac{\mathrm{d}}{\mathrm{d}\psi}\,P(\psi' \leq \psi \mid E).
\end{equation}
For track events, the PSF improves rapidly with energy: in the IceCube point-source sample~\citep{IceCube:2021xar} the median angular error is $1$--$1.5^\circ$ at \SI{1}{\TeV}, falling to $\sim 0.5^\circ$ above \SI{10}{\TeV}.
For cascade events the median is an order of magnitude larger, limited by the short lever arm of the shower rather than by photon statistics.

The PSF is stored as an inverse CDF $\psi(E, u)$ over a grid of true energies and uniform quantiles; see \cref{app:hdf5} for further details.
A reconstructed direction is obtained by drawing a deflection angle $\psi$ from the inverse CDF at the event's true energy, then rotating the true direction by $(\psi, \phi)$ where $\phi \sim \mathrm{Uniform}(0, 2\pi)$.

\subsection{Energy Resolution}
\label{sec:irf:eres}

The energy resolution characterizes the smearing between true neutrino energy $E_{\nu}$ and reconstructed deposited energy $E_\mathrm{reco}$.
The energy resolution is stored as an inverse CDF of the log-ratio $\Delta \equiv \ln(E_\mathrm{reco}/E_\mathrm{true})$, so a reconstructed energy is obtained as $E_\mathrm{reco} = E_\mathrm{true} \cdot e^{\Delta}$ where $\Delta$ is drawn from the stored quantile function.

In general, cascade events have better energy resolution because most or all photons remain within the detector fiducial volume.
Tracks, on the other hand, may enter and leave the detector (through-going), start in the detector but leave (starting tracks), enter the detector from outside but stop (stopping tracks), or start and stop in the detector (contained tracks).
This last class, where calorimetry might be possible, is quite rare in many event selections as muons can travel several kilometers in ice at \SI{1}{\TeV}~\citep{Koehne:2013gpa,Dunsch:2018nsc}.
Due to this, energies are typically estimated by measuring the muon energy loss profile, $\frac{\mathrm{d}E}{\mathrm{d}x}$, which scales linearly with the muon energy in this regime~\citep{Groom:2001kq}.
This is typically less precise than calorimetric measurements of energy.

For IRF files that encode the joint distribution $P(E_\mathrm{reco}, \psi, \sigma_\psi \mid E_\mathrm{true}, \delta)$ as a five-dimensional histogram---the format used in the IceCube ten-year point-source data release---a single draw simultaneously returns the reconstructed energy, deflection angle, and per-event angular uncertainty $\sigma_\psi$.

\section{Source Models}
\label{sec:sources}

All source models in \spore{} specify the differential neutrino flux $\mathrm{d}N/(\mathrm{d}E\, \mathrm{d}A\,\mathrm{d}t\,\mathrm{d}\Omega)$, in units of \si{\GeV^{-1} \cm^{-2} \s^{-1} sr^{-1}}, as a function of neutrino species and energy.
The sampler classes in \cref{sec:algorithms} convolve this flux with the detector response to produce Monte Carlo event samples, and are agnostic to the specific source geometry or spectral shape in use.

\spore{} models sources as either point-like or extended.
A point source is defined by a fixed sky coordinate (declination and right ascension) and a differential flux $\Phi^{1\nu}(E_{\nu})$ for each of the six neutrino species.
The source itself is detector-independent; the conversion from flux to expected event rate is handled entirely by the sampler at sampling time.
An extended source, on the other hand, specifies the differential flux per steradian as a function of direction and energy.
This can be used to model any flux that has spatial extent.
For example, the flux of atmospheric neutrinos is treated in this way.
Furthermore, neutrinos from the Galactic plane or from dark matter annihilation in the Galactic halo may also be implemented with this method.

All source fluxes encode a per-species normalization and a spectral (and optionally spatial) shape.
Four built-in implementations are provided for standard use cases: an analytic power law, and tabulated fluxes in one, two, or three dimensions.
The power law is
\begin{equation}
  \Phi^{1\nu}(E_{\nu}) = \begin{cases}
      0  & E_{\nu} < E_{\mathrm{min}}\\
      \Phi_0^{1\nu} \times
  \left(\frac{E_{\nu}}{E_\mathrm{pivot}}\right)^{-\gamma} & E_{\mathrm{min}} \leq E_{\nu} \leq E_{\mathrm{max}}\\
    0 & E_{\mathrm{max}} < E_{\nu}
  \end{cases},
  \label{eq:powerlaw}
\end{equation}
parameterized by single-species normalization $\Phi_0^{1\nu}$, spectral index $\gamma$, pivot energy $E_\mathrm{pivot}$, and hard cutoffs $E_\mathrm{min}$ and $E_\mathrm{max}$.

\spore{} also supports user-supplied spectra in the form of tabulated fluxes.
These can be only energy-dependent, in which case the flux is assumed to be isotropic; energy- and declination-dependent, in which case the flux is assumed to be uniform in right ascension; or energy-, declination-, and right-ascension-dependent.
\spore{} infers which case to use from the dimensionality of the input flux.
This allows models from external calculators (e.g., atmospheric neutrino flux tables from \textsc{nuflux}~\citep{nuflux} or MCEq~\citep{Fedynitch:2015zma,Fedynitch:2018cbl}) to be ingested directly.

All flux classes support construction from TOML configuration files; the format is described in \cref{app:toml}.

\section{Event Sampling Algorithms}
\label{sec:algorithms}

Two sampler classes cover the principal source categories encountered in high-energy neutrino astronomy; both accept a list of detectors directly, so multi-detector joint analyses require no third class.
All samplers precompute inverse-CDF tables at construction time and share a common interface for generating events and computing expected counts.
The observation livetime $\Delta t$ is passed only at sampling time, while the choice of transient versus steady-state effective-area averaging is fixed at construction.
PSF and energy smearing are applied identically in both classes via a shared utility, so any improvement to the smearing logic propagates automatically.

\subsection{Point Source Sampler}
\label{sec:algorithms:ps}

The point-source sampler generates reconstructed neutrino events from an astrophysical point source specified by a sky coordinate and a differential flux model.

Given the source location and a reference epoch, the expected differential event rate is
\begin{equation}
  \frac{\mathrm{d}N}{\mathrm{d}E_{\nu}\,\mathrm{d}t\,\mathrm{d}\Omega}(\zeta,\, E_{\nu}) =
  A_{\mathrm{eff}}(\zeta,\, E_{\nu})\,\Phi(E_{\nu})
\end{equation}
where $A_{\mathrm{eff}}(\zeta, E_{\nu})$ is the morphology-specific effective area and $\Phi$ sums over the neutrino flavors relevant to the requested morphology: $\nu_\mu + \bar\nu_\mu$ for track-like events, and all six species for cascade-like events.
The cumulative distribution function of the energy spectrum,
\begin{equation}
  C(\ln E) = \int_{\ln E_{\min}}^{\ln E}
              A_{\mathrm{eff}}\!\left(\zeta, e^{u}\right)
              \Phi\!\left(e^{u}\right)
              e^{u}\, \mathrm{d}u,
\end{equation}
is computed by numerical quadrature over $\ln E$.
The normalization $\lambda = C(\ln E_{\max})$ gives the expected total event rate in units of \si{\s^{-1}}.
True energies are sampled by drawing uniform variates and evaluating the inverse CDF, and the true arrival direction is the source sky coordinate.
Each event is then smeared by the PSF and energy resolution.

\subsection{Extended Source Sampler}
\label{sec:algorithms:extended}

The extended source sampler handles sources with a non-trivial angular distribution on the sky, including Galactic plane emission templates, diffuse emission from star-forming regions, or any user-supplied flux map.
Because the effective area depends on zenith angle, which varies continuously across the sky, the joint distribution $p(\delta, \alpha, E)$ cannot, in general, be factored analytically.
\spore{} uses a fully discrete hierarchical inverse-CDF approach.
At construction time, the sampler precomputes the target weight function
\begin{equation}
  w(i, j, k) =
  A_{\mathrm{eff}}\!\left(\zeta_{ij},\, E_k\right)
  \times
  \Phi\!\left(\delta_i,\, \alpha_j,\, E_k\right)
  \label{eq:target_weight}
\end{equation}
on a three-dimensional grid of $n_{\delta}$ uniformly spaced $\sin\delta$ bins, $n_\alpha$ uniformly spaced right-ascension bins, and $n_E$ energy bins (default $n_\delta = n_\alpha = n_E = 40$).
The energy range is taken from the intersection of the support of the effective area with that of the flux model, unless the user overrides it.
The energy grid is adaptive: cell centres are placed at equal quantiles of the integrated weight $A_\mathrm{eff}(E)\,\Phi(E)\,E\,\mathrm{d}\ln E$, which concentrates resolution near the detection threshold where $A_\mathrm{eff}$ rises steeply.
Because the sampler draws $\ln E$ uniformly within a cell, pure equal-mass placement is unsafe at the depleted end of the range: for a steeply falling atmospheric spectrum nearly all of the weight lies below a few tens of TeV, so the final quantile step can span more than a decade and smear its entire mass flat across that decade.
The quantile placement is therefore blended with a uniform-in-$\ln E$ one, which bounds the widest cell at twice the uniform spacing while retaining most of the resolution near threshold.
For azimuthally symmetric sources the flux is independent of $\alpha_j$ and the sampler broadcasts $\Phi(\delta_i, E_k)$ uniformly across the RA axis; for RA-dependent sources it is evaluated at every $(\delta_i, \alpha_j)$ cell.

\begin{figure}[t]
  \centering
  \includegraphics[width=\textwidth]{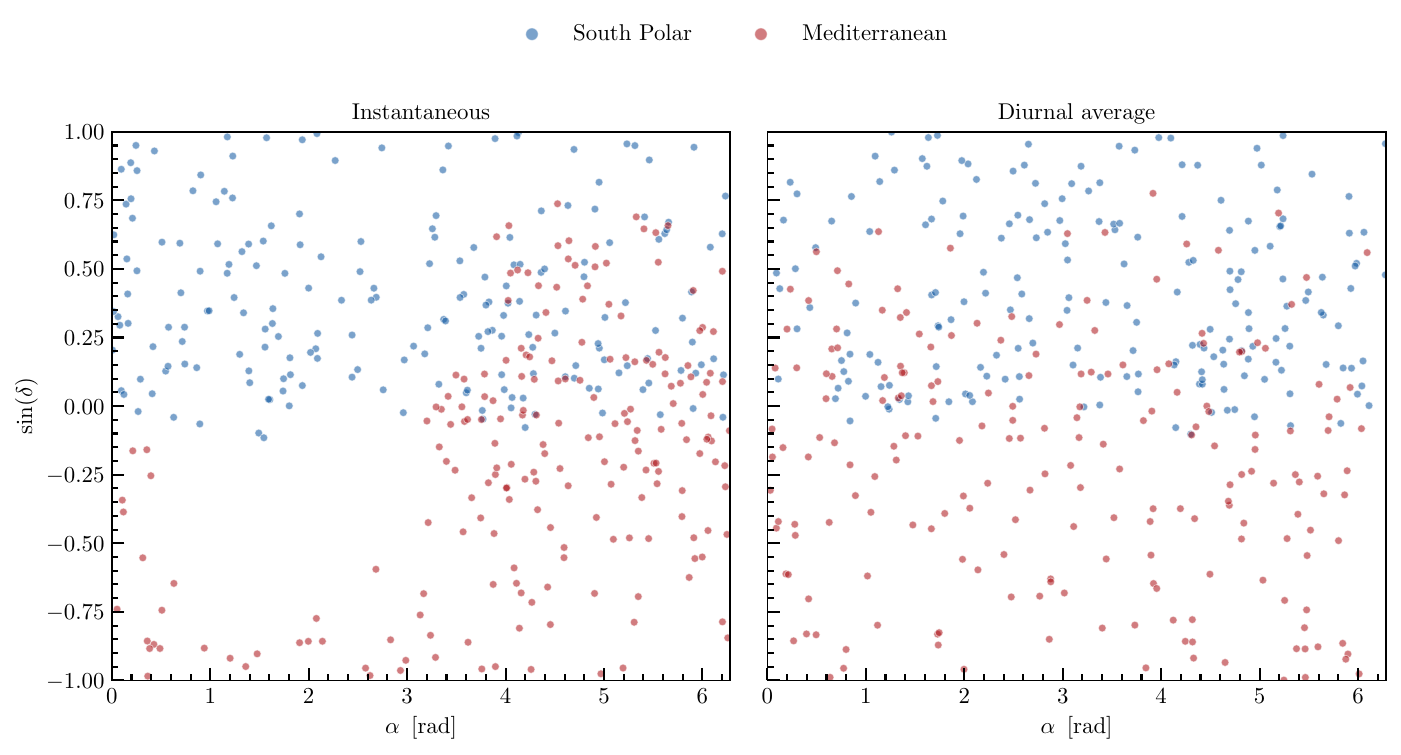}
  \caption{
    \textbf{\textit{Reconstructed event directions for Mediterranean and South Polar neutrino telescopes.}}
    The left panel shows the instantaneous exposure, while the right panel shows the diurnally averaged exposure.
    The distribution from the South Polar detector (blue) does not change from one panel to the next due to its position on Earth's rotation axis, whereas the Mediterranean detector (red) is smeared by the diurnal averaging.
  }
  \label{fig:multidet_skymap}
\end{figure}

From the weight grid $w(i,j,k)$ the sampler precomputes CDFs that factor the joint distribution as
\begin{equation}
  p(\sin\delta,\,\alpha,\,\ln E)
  =
  p(\sin\delta)
  \;\times\;
  p(\ln E \mid \sin\delta)
  \;\times\;
  p(\alpha \mid \sin\delta,\,\ln E).
  \label{eq:factored}
\end{equation}
To draw $N_\mathrm{event}$ events, three uniform variates are drawn per event and passed through the corresponding inverse CDFs to select grid bin indices; each coordinate is then jittered uniformly within its cell to avoid aliasing.
While this design choice introduces an error that is linear in the bin size, it handles multimodal distributions---for example, the angular acceptance gap between upgoing and downgoing track events at intermediate zenith angles---without any MCMC burn-in.

\subsection{Multi-Detector Joint Analyses}
\label{sec:algorithms:multidet}

Multi-detector joint simulations require no additional configuration beyond passing multiple detectors to either sampler class.
Each detector is treated independently: it contributes its own effective area, PSF, and energy resolution, and its geographic location determines the zenith angle at which the source is observed.

Each detector independently Poisson-samples its expected event count:
\begin{equation}
  N_d \sim \mathrm{Poisson}\!\left(\lambda_d \,\Delta t_d\right),
  \qquad
  \lambda_d = \int A_\mathrm{eff}^{(d)}(\zeta_d, E)\,
              \Phi(E)\,\mathrm{d}E,
\end{equation}
where the detector-specific zenith angle $\zeta_d$ reflects the source position as seen from detector $d$'s location at the observation epoch, via \cref{eq:zenith}.
The observation livetime and reference epoch can be supplied as shared values or specified independently per detector.
Each event carries an integer detector index, so downstream likelihood analyses can trivially separate or combine events from different instruments while sharing a common source flux normalization.
Full API details are given in \cref{app:architecture}.

\subsection{Sampling Modes}

When sampling events, the user may ask for events in two modes: \emph{Poisson mode} and \emph{fixed-$N$ mode}.
In Poisson mode, the number of events is drawn as $N \sim \mathrm{Poisson}(\lambda\,\Delta t)$, where $\Delta t$ is the observation livetime passed as an argument.
This properly accounts for the flux normalization.
In fixed-$N$ mode, exactly $N$ events are generated, useful for constructing ensemble test statistics.

Additionally, the user may specify whether they want the sampler to be constructed in \emph{instantaneous} or \emph{steady-state} mode.
In instantaneous mode, the effective area is evaluated at a single reference epoch, appropriate for short-duration observations, such as those used for transient phenomena.
In steady-state mode, the effective area is averaged over a full diurnal cycle:
\begin{equation}
  \langle A_{\mathrm{eff}}(\delta, E) \rangle
  =
  \frac{1}{2\pi}
  \int_0^{2\pi}
  A_{\mathrm{eff}}\!\left(\zeta(\delta, H),\, E\right) \mathrm{d}H,
  \label{eq:ha_avg}
\end{equation}
with $\zeta(\delta, H)$ given by \cref{eq:zenith}.
In practice the integral is evaluated as a sum over a user-specifiable number of uniformly spaced sidereal phases (default 100).
\cref{fig:multidet_skymap} illustrates the difference between the two modes for a polar and a mid-latitude detector observing the same sky: the South Polar sample is unchanged by the diurnal average, while the Mediterranean one is smeared across right ascension.
Each phase carries its own instantaneous sampling grid, and an event is assigned a phase in proportion to the rate there, so that its sky position, energy, and arrival time all correspond to the same observing configuration.
Steady-state mode is appropriate for analyses spanning many sidereal days.
If a livetime exceeding 15~minutes is passed to a transient-mode sampler, a warning is issued to alert the user that the single-epoch effective area may underestimate the true time-averaged rate for non-polar detectors.


\section{Validation}
\label{sec:validation}

We validate the sampling framework via three complementary tests.

\subsection{HESE 7.5-year validation}
\label{sec:validation:hese}

In order to validate the sampling procedure, we compare predicted event rates and energy spectra against the IceCube High-Energy Starting Event (HESE) 7.5-year public data release~\citep{IceCube:2020wum}.
The HESE sample selects events with deposited energy above $\sim$60~TeV that begin inside the detector, providing a clean sample of astrophysical neutrinos with negligible atmospheric muon background.
We model the expected event counts following the best-fit parameters from Ref.~\citep{IceCube:2020wum}: an isotropic astrophysical power-law flux with spectral index $\gamma = 2.87$ and a per-species normalization $\Phi_0^{1\nu} = \SI{1.06e-18}{\GeV^{-1}\cm^{-2}\s^{-1}\steradian^{-1}}$ at a pivot energy of \SI{100}{\TeV}, one sixth of the all-flavor total, plus a conventional atmospheric component at its nominal normalization, $\Phi_\mathrm{conv} = 1.00$, with cosmic-ray tilt $\Delta\gamma_\mathrm{CR} = -0.053$, kaon-to-pion ratio $1.0001$, and $\nu/\bar\nu$ ratio $0.998$.

\begin{figure}[t]
  \centering
  \includegraphics[width=\textwidth]{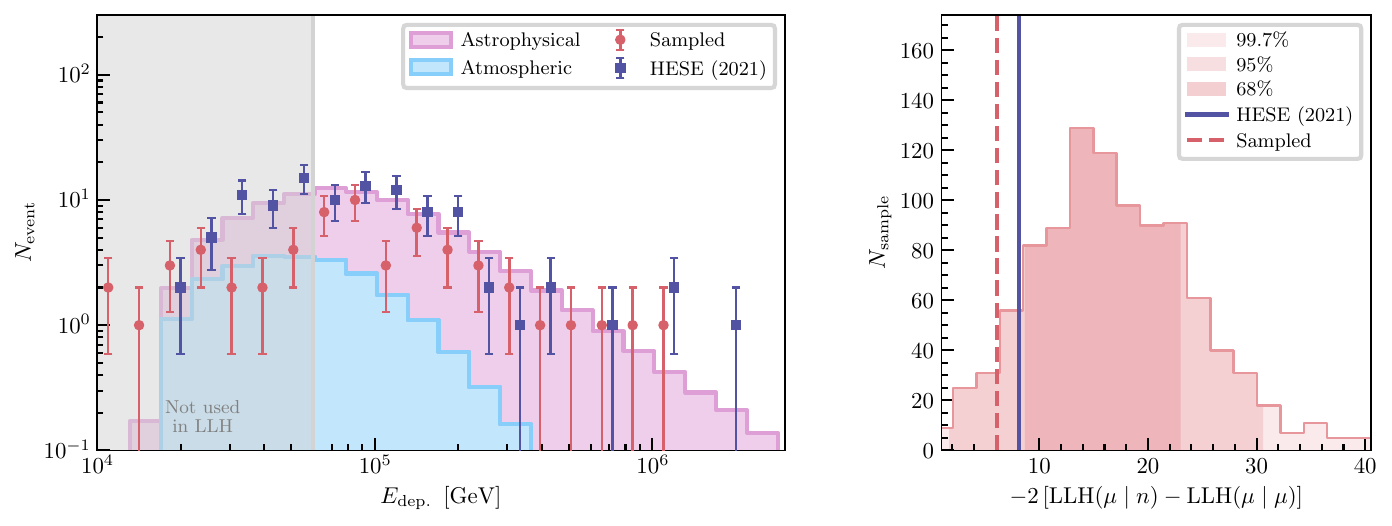}
  \caption{
    \textbf{\textit{Sampled events compared to experimental observation and MC expectation from HESE 7.5-yr.}}
    The left panel shows stacked filled regions for the atmospheric (blue) and astrophysical (pink) MC expectations at the best-fit parameters from Ref.~\citep{IceCube:2020wum} with nominal detector systematics.
    Red points show a single draw of the \spore{} sampling, while dark blue points show the observed experimental data.
    The gray band marks the region below \SI{60}{\TeV} where atmospheric muons still contribute, which \spore{} does not simulate and which the HESE analysis excludes.
    The right panel shows the distribution of test statistics between 1{,}000 \spore{} samples and the best-fit Monte Carlo.
    The shaded regions denote the minimum-length 68\%, 95\%, and 99.7\% containment regions and the test statistics from the \spore{} sample in the left panel and the experimental data are shown as vertical lines.
  }
  \label{fig:hese_energy}
\end{figure}

We use the publicly available Monte Carlo~\citep{IceCube:2020wum} to create the requisite IRF and set the livetime to $T = 2{,}635$~days.
The left panel of \cref{fig:hese_energy} compares the predicted stacked spectra of astrophysical and atmospheric components from Monte Carlo to a single \spore{} sample generated with this IRF and to the 102 observed HESE events.
The gray box denotes the region below \SI{60}{\TeV} where the HESE analysis cuts to avoid atmospheric muon contamination.

The right panel of the same figure shows the distribution of test statistics obtained from drawing 1{,}000 \spore{} samples and computing twice the log-likelihood difference between the likelihood from comparing sampled events to the best-fit Monte Carlo and comparing the Monte Carlo to itself.
Since \spore{} does not model atmospheric muons, we follow the HESE prescription of only using events with deposited energy above \SI{60}{\TeV}.
The likelihood values from the sample on the right and the observed experimental events are shown as vertical lines.
The test statistic of the observed data lies within the range spanned by the sampled pseudo-experiments, indicating that \spore{} reproduces both the mean expectation and the size of the Poisson scatter about it.
It does not, however, lie in the middle of that range: only $11.3\%$ of the sampled realizations agree with the best-fit Monte Carlo more closely than the observed events do.
This is expected rather than anomalous.  The Monte Carlo expectation is the published best fit to these same data, so the observed counts sit closer to it than an independent Poisson realization typically would.
The comparison therefore tests whether the sampler produces data-like fluctuations about a given expectation; it is not an independent test of the underlying flux model.


\subsection{Round-trip consistency: IceCube 10-year tracks}
\label{sec:validation:roundtrip}

An instrument response is built from the publicly available IceCube 10-year point-source data release \citep{IceCube:2021xar} and used to predict the declination and reconstructed-energy distributions of northern-sky ($\delta > 0$) track events, which are then compared to the observed IC86 sample (seasons IC86-I to IC86-VII, \num{2531.9}~days of livetime).

The expected event rate is modeled as the sum of an astrophysical component and a conventional atmospheric component.
The astrophysical flux is taken from a tabulated model derived from the 10-year point-source analysis~\citep{IceCube:2021xar}.
The atmospheric flux is modeled using MCEq~\citep{Fedynitch:2015zma,Fedynitch:2018cbl} with the Global Spline Fit cosmic-ray flux model~\citep{Dembinski:2017zsh} and the Sibyll~2.3d hadronic interaction model~\citep{Riehn:2019jet}.
The overall atmospheric normalization is fit to the observed data by maximizing a Poisson log-likelihood, yielding a fitted normalization of 1.30 relative to the nominal MCEq prediction.
This offset is consistent with the known O(20--30\%) absolute normalization uncertainty on uncalibrated conventional atmospheric flux predictions in this energy range~\citep{Yanez:2023lsy}.
All distributions are produced by averaging 40 \spore{} pseudo-experiments drawn from an extended-source sampler configured for a South Pole detector and the full observed livetime.

\begin{figure}[t]
  \centering
    \includegraphics[width=\textwidth]{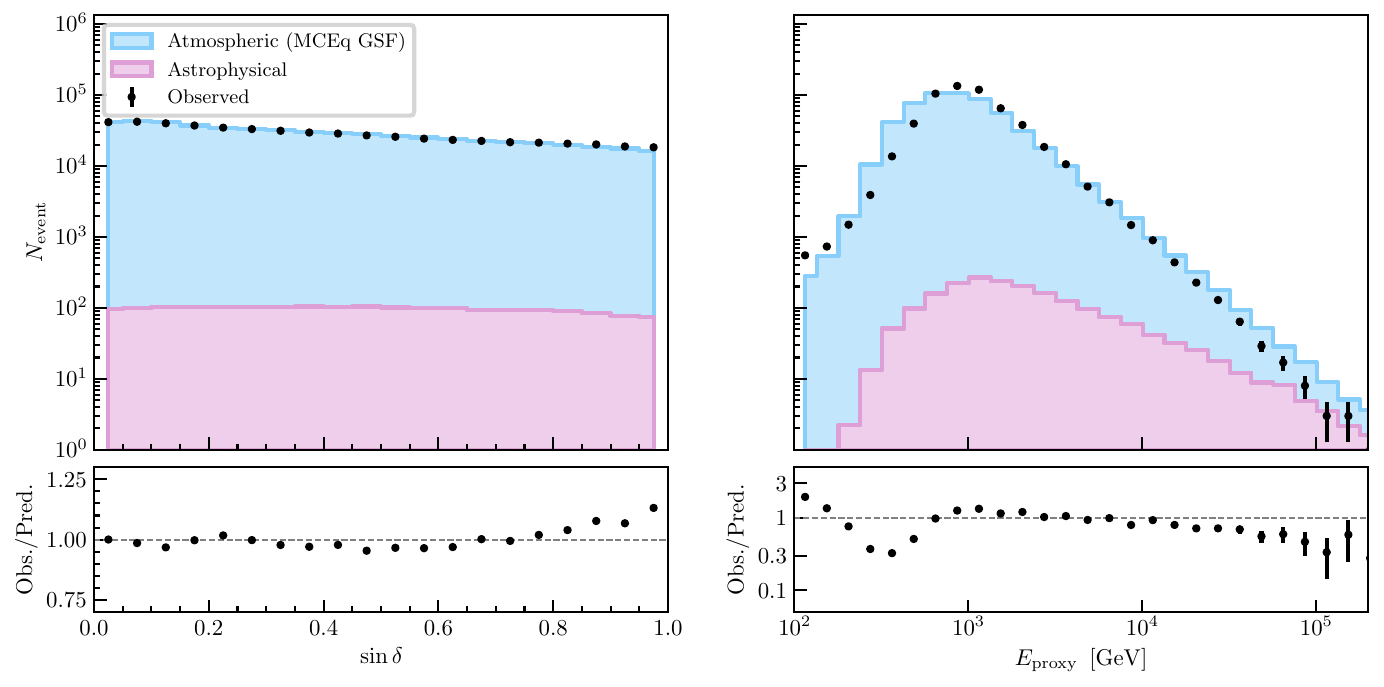}
    \caption{
    \textbf{\textit{Sampling expectation for northern-sky tracks versus observed data.}}
    The upper panels show the predictions from 40 \spore{} pseudo-experiments using the IceCube 10-year point-source IRF~\citep{IceCube:2021xar}, decomposed into atmospheric (blue, MCEq GSF) and astrophysical (pink) components, compared to the observed IC86 data (black points).
    The left column shows the distribution in $\sin\delta$ and the right column the distribution in reconstructed muon energy proxy, both for northern-sky ($\delta > 0$) events.
    The atmospheric normalization is fit to the $\sin\delta$ distribution only, so the energy proxy distribution is a prediction.
    The lower panels show the ratio of observed to predicted counts; the dashed line marks unity.
    Note that the residual axis is linear on the left and logarithmic on the right.
    }
  \label{fig:roundtrip_energy}
\end{figure}

The left panels of \cref{fig:roundtrip_energy} show the stacked atmospheric (blue) and astrophysical (pink) predictions alongside the observed northern-sky track data as a function of $\sin\delta$.
The ratio panel shows agreement at the 10--15\% level across the full declination range.  A residual tilt remains, with the prediction falling slightly more steeply from horizon to pole than the data; we have not identified its origin, though declination-dependent detector systematics not encoded in the public effective area tables would produce an effect of this kind.

The right panels repeat the comparison as a function of the reconstructed muon energy proxy, applying the same northern-sky selection.
This is a considerably more stringent test than the declination distribution: the atmospheric normalization is fit to the $\sin\delta$ distribution alone, so the energy proxy shape is a genuine prediction, and it exercises the joint smearing matrix rather than only the effective area.
The prediction reproduces the shape of the observed distribution over the bulk of the sample, agreeing to within about a third from the peak of the distribution, near \SI{800}{\GeV}, up to a few tens of TeV.

Below $\sim\SI{600}{\GeV}$, however, the prediction exceeds the data by up to a factor of three.
We attribute this to the granularity of the public smearing matrix rather than to the sampling itself.
The released tables provide $P(E_\mathrm{proxy} \mid E_\nu, \delta)$ in true-energy bins of width $0.5$ in $\log_{10}(E_\nu/\mathrm{GeV})$ and in only three declination bands, and the kernel within each bin is an average weighted by the simulation spectrum used to produce the release rather than by the physical atmospheric spectrum.
For a spectrum falling as steeply as the conventional atmospheric flux, this bin-averaging biases the kernel toward the upper edge of each true-energy bin and broadens it, which populates the low-energy proxy region where the event selection is in fact turning on.
The effect is intrinsic to the released IRFs: repeating the calculation as a direct forward fold of the effective area and smearing matrix, with no event sampling at all, over-predicts the observed rate near \SI{400}{\GeV} by a comparable factor of $\sim$2.
A comparable treatment would require the unbinned resolution functions available internally to the collaboration.

\subsection{Point source analysis: angular distribution}
\label{sec:validation:ps}

As a third validation, we demonstrate the $\psi^2$ analysis workflow used in IceCube point source searches~\citep{Braun:2008bg,IceCube:2019cia}.
A synthetic event sample is generated with the IceCube 10-year point-source IRF, combining an astrophysical background, an atmospheric background, and a signal contribution at the location of NGC-1068 with spectral parameters taken from Ref.~\citep{IceCube:2022der}.
These were sampled with a livetime of $3{,}186$~days taken from the same reference.

\begin{figure}[t]
  \centering
  \includegraphics[width=0.7\textwidth]{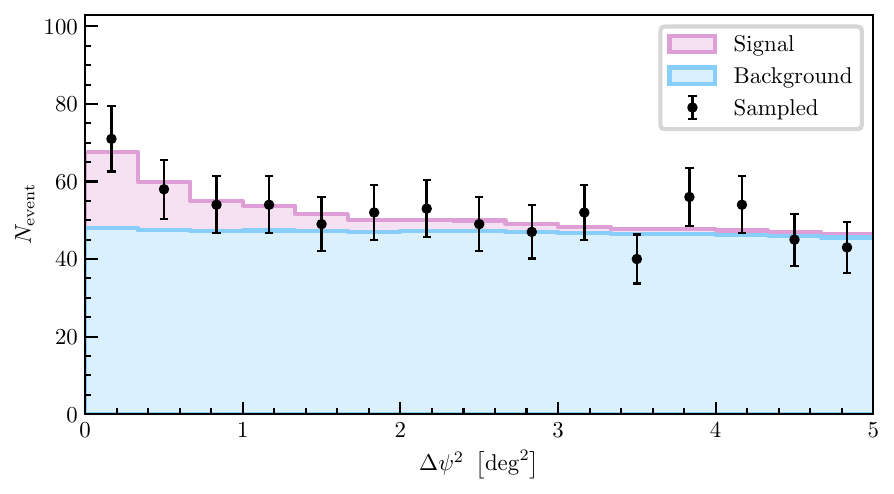}
  \caption{
    \textbf{\textit{Angular distribution of events near NGC-1068.}}
    The signal template (pink), constructed by averaging 1{,}000 point-source pseudo-experiments, and the background template (blue), constructed by averaging 1{,}000 RA-scrambled realizations of the synthetic event set, are shown separately.
    The point source is placed at the location of NGC-1068 with $\gamma = 3.2$ and a per-species normalization $\Phi^{1\nu}_{1\,\mathrm{TeV}} = \SI{2.5e-14}{\GeV^{-1}\cm^{-2}\s^{-1}}$, so that the summed $\nu_\mu + \bar\nu_\mu$ flux sampled for track events matches the best fit of Ref.~\citep{IceCube:2022der}.
    The $\psi^2$ distribution of the synthetic event sample (atmospheric + astrophysical background + NGC-1068 signal) is shown as black scatter points; the excess concentrated near $\psi^2 = 0$ reflects the signal contribution.
    The realization shown is one of 60 draws, chosen to be free of large bin-to-bin fluctuations; the templates and all quantitative statements use the full unselected ensembles.
  }
  \label{fig:ps_psi2}
\end{figure}

The background template is built using the standard right-ascension scrambling technique~\citep{Braun:2008bg}: the synthetic event set is drawn 1{,}000 times with right-ascension coordinates replaced by independent uniform draws on $[0, 2\pi)$, preserving the declination distribution while erasing any source-correlated clustering.
The mean $\psi^2$ histogram over these scrambled realizations defines the expected background shape.
A signal template is constructed by independently sampling 1{,}000 Poisson pseudo-experiments from the point source alone and averaging the resulting $\psi^2$ distributions.

\cref{fig:ps_psi2} compares these templates to the $\psi^2$ histogram of the synthetic event sample.
The background template matches the synthetic data distribution at large $\psi^2$ where signal is negligible, and the signal template shows the expected excess concentrated near $\psi^2 = 0$.
This exercises both the PSF sampling, which sets the shape of the signal cone, and the declination-dependent effective-area weighting, which sets the normalization of the background template.

The shape and normalization are approximately consistent with the data shown in Ref.~\citep{IceCube:2022der}.  The signal template here is slightly wider; reconstruction improvements made since the ten-year data release would account for a difference in this direction, but we have not directly verified that this is the cause.

\section{Limitations}
\label{sec:limitations}

Several limitations are worth stating directly.
The sampled distributions are exact only up to the resolution of the sampling grid: coordinates are jittered uniformly within their cells, so the error is linear in cell size, and the default $40 \times 40 \times 40$ grid should be increased for sources with structure finer than a cell.
The fidelity of the output is bounded by the fidelity of the input IRFs, and the public data releases used here are coarse: the ten-year point-source smearing matrix is binned in true-energy steps of $0.5$ in $\log_{10} E$ and in three declination bands, with each kernel averaged over the simulation spectrum used to produce the release rather than the physical spectrum being sampled.
This is the dominant systematic in the energy round-trip of \cref{sec:validation:roundtrip} and cannot be removed downstream.
The effective-area smoothing and trimming described in \cref{app:hdf5} are applied by default and modify the tabulated response before sampling.
Neutrino absorption in the Earth, flavor composition changes en route, and detector systematic uncertainties are not modeled separately; they are present only insofar as they are already folded into the tabulated IRFs.
At present, the \texttt{medium} field accepted in the detector configuration is recorded but never used, and \texttt{depth} is ignored entirely.
Atmospheric muons are not simulated, and no IRF for them is distributed.
Finally, morphology labels are supplied by the response file rather than derived from interaction physics: for a single-species effective area, track morphologies are assigned the $\nu_\mu + \bar\nu_\mu$ flux and all others the six-species sum, which neglects the small track contribution from $\nu_\tau$ charged-current interactions.
Response files that resolve all six species separately avoid this approximation.

\section{Conclusions}
\label{sec:conclusions}

We have presented \spore{}, an open-source Python package for generating realistic simulated neutrino events from astrophysical sources using tabulated instrument response functions.
The package provides two sampler classes---point source and extended source---sharing a common hierarchical inverse-CDF engine that avoids MCMC burn-in and whose discretization error is linear in the cell size; both natively support multi-detector joint simulations without requiring a separate class or configuration.
Fully RA-dependent flux models are handled natively by the extended source sampler: loading a three-dimensional tabulated flux automatically enables full $(\sin\delta, \alpha, E)$ grid evaluation without any additional user configuration.

The IRF-agnostic HDF5 format decouples the simulation machinery from any particular detector or analysis selection.
Any neutrino telescope for which an effective area, PSF, and energy resolution are available---whether from a public data release or a private study---can be ingested into the package without code modification.
We validated the framework against the IceCube 10-year point-source and HESE 7.5-year public data releases: the northern-sky declination distribution is reproduced at the 10--15\% level, the sampled HESE deposited-energy spectrum tracks the published expectation, and the $\psi^2$ background shape of a standard point-source search is recovered.
The reconstructed-energy round trip agrees over the bulk of the sample but not below $\sim\SI{600}{\GeV}$, for the reasons given in \cref{sec:validation:roundtrip}.

\spore{} fills a gap in the existing neutrino simulation ecosystem.  We are not aware of another publicly available tool that (i) produces final-level reconstructed events (direction, energy, time, morphology, and local zenith and azimuth) rather than detector-level simulation output, (ii) supports a broad range of source hypotheses within a single framework, and (iii) extends to a new observatory---IceCube-Gen2, KM3NeT/ARCA, or P-ONE---by supplying a response file rather than by modifying code.  Detectors whose response is not well described by an effective area together with angular and energy resolutions, such as radio arrays, would require the response format to be extended.

Future development will focus on expanding the library of ready-to-use IRF files distributed with the package, adding support for time-dependent effective areas (e.g.\ for detectors still under construction), and providing a higher-level sensitivity-scan interface that automates pseudo-experiment loops over source normalizations and spectral indices.
Contributions are welcome via the public repository.

\section*{Code availability}

\spore{} is openly available at \url{https://github.com/jlazar17/spore} under the LGPL-3.0-or-later license.
The repository includes the instrument response files for the two IceCube data releases used here, together with the scripts that generate every figure in this paper; the public data releases themselves must be downloaded from their respective archives, as described in the repository documentation.

\acknowledgments

J.L. and G.d.W. acknowledge support from the Fonds de la Recherche Scientifique (F.R.S.--FNRS).

\bibliographystyle{elsarticle-num}
\bibliography{references}

\begin{thebibliography}{10}
\expandafter\ifx\csname url\endcsname\relax
  \def\url#1{\texttt{#1}}\fi
\expandafter\ifx\csname urlprefix\endcsname\relax\def\urlprefix{URL }\fi
\expandafter\ifx\csname href\endcsname\relax
  \def\href#1#2{#2} \def\path#1{#1}\fi

\bibitem{Meszaros:2019xej}
P.~Mész{\'a}ros, D.~B. Fox, C.~Hanna, K.~Murase, {Multi-Messenger
  Astrophysics}, Nature Rev. Phys. 1 (2019) 585--599.
\newblock \href {http://arxiv.org/abs/1906.10212} {\path{arXiv:1906.10212}},
  \href {https://doi.org/10.1038/s42254-019-0101-z}
  {\path{doi:10.1038/s42254-019-0101-z}}.

\bibitem{IceCube:2013low}
M.~G. Aartsen, et~al., {Evidence for High-Energy Extraterrestrial Neutrinos at
  the IceCube Detector}, Science 342 (2013) 1242856.
\newblock \href {http://arxiv.org/abs/1311.5238} {\path{arXiv:1311.5238}},
  \href {https://doi.org/10.1126/science.1242856}
  {\path{doi:10.1126/science.1242856}}.

\bibitem{IceCube:2020wum}
R.~Abbasi, et~al., {The IceCube high-energy starting event sample: Description
  and flux characterization with 7.5 years of data}, Phys. Rev. D 104 (2021)
  022002.
\newblock \href {http://arxiv.org/abs/2011.03545} {\path{arXiv:2011.03545}},
  \href {https://doi.org/10.1103/PhysRevD.104.022002}
  {\path{doi:10.1103/PhysRevD.104.022002}}.

\bibitem{IceCube:2025tgp}
R.~Abbasi, et~al., {Evidence for a Spectral Break or Curvature in the Spectrum
  of Astrophysical Neutrinos from 5 TeV--10 PeV}, Phys. Rev. Lett. 136 (2026)
  121002.
\newblock \href {http://arxiv.org/abs/2507.22233} {\path{arXiv:2507.22233}},
  \href {https://doi.org/10.1103/2gh9-d4q7} {\path{doi:10.1103/2gh9-d4q7}}.

\bibitem{IceCube:2018cha}
M.~G. Aartsen, et~al., {Neutrino emission from the direction of the blazar TXS
  0506+056 prior to the IceCube-170922A alert}, Science 361~(6398) (2018)
  147--151.
\newblock \href {http://arxiv.org/abs/1807.08794} {\path{arXiv:1807.08794}},
  \href {https://doi.org/10.1126/science.aat2890}
  {\path{doi:10.1126/science.aat2890}}.

\bibitem{IceCube:2022der}
R.~Abbasi, et~al., {Evidence for neutrino emission from the nearby active
  galaxy NGC 1068}, Science 378 (2022) 538--543.
\newblock \href {http://arxiv.org/abs/2211.09972} {\path{arXiv:2211.09972}},
  \href {https://doi.org/10.1126/science.abg3395}
  {\path{doi:10.1126/science.abg3395}}.

\bibitem{KM3Net:2016zxf}
S.~Adrian-Martinez, et~al., {Letter of intent for KM3NeT 2.0}, J. Phys. G
  43~(8) (2016) 084001.
\newblock \href {http://arxiv.org/abs/1601.07459} {\path{arXiv:1601.07459}},
  \href {https://doi.org/10.1088/0954-3899/43/8/084001}
  {\path{doi:10.1088/0954-3899/43/8/084001}}.

\bibitem{P-ONE:2020ljt}
M.~Agostini, et~al., {The Pacific Ocean Neutrino Experiment}, Nature Astron. 4
  (2020) 913--915.
\newblock \href {http://arxiv.org/abs/2005.09493} {\path{arXiv:2005.09493}},
  \href {https://doi.org/10.1038/s41550-020-1182-4}
  {\path{doi:10.1038/s41550-020-1182-4}}.

\bibitem{TRIDENT:2022hql}
Z.~P. Ye, et~al., {A multi-cubic-kilometre neutrino telescope in the western
  Pacific Ocean}, Nature Astron. 7 (2023) 1497--1505.
\newblock \href {http://arxiv.org/abs/2207.04519} {\path{arXiv:2207.04519}},
  \href {https://doi.org/10.1038/s41550-023-02087-6}
  {\path{doi:10.1038/s41550-023-02087-6}}.

\bibitem{IceCube-Gen2:2020qha}
M.~G. Aartsen, et~al., {IceCube-Gen2: the window to the extreme Universe}, J.
  Phys. G 48 (2021) 060501.
\newblock \href {http://arxiv.org/abs/2008.04323} {\path{arXiv:2008.04323}},
  \href {https://doi.org/10.1088/1361-6471/abbd48}
  {\path{doi:10.1088/1361-6471/abbd48}}.

\bibitem{Thompson:2023pnl}
W.~G. Thompson, {TAMBO: Searching for Tau Neutrinos in the Peruvian Andes}, in:
  {38th International Cosmic Ray Conference}, 2023.
\newblock \href {http://arxiv.org/abs/2308.09753} {\path{arXiv:2308.09753}}.

\bibitem{Arguelles:2026btb}
C.~A. Arg\"uelles, et~al., {Measuring the high-energy neutrino sky using the
  deep-valley neutrino observatory TAMBO}, Nature Astron. 10~(7) (2026)
  947--951.
\newblock \href {http://arxiv.org/abs/2507.08070} {\path{arXiv:2507.08070}},
  \href {https://doi.org/10.1038/s41550-026-02916-4}
  {\path{doi:10.1038/s41550-026-02916-4}}.

\bibitem{Otte:2019knb}
A.~N. Otte, et~al., {Trinity: An Air-Shower Imaging System for the Detection of
  Ultrahigh Energy Neutrinos}, PoS ICRC2019 (2020) 976.
\newblock \href {http://arxiv.org/abs/1907.08732} {\path{arXiv:1907.08732}},
  \href {https://doi.org/10.22323/1.358.0976} {\path{doi:10.22323/1.358.0976}}.

\bibitem{GRAND:2018iaj}
J.~{\'A}lvarez-Mu{\~n}iz, et~al., {The Giant Radio Array for Neutrino Detection
  (GRAND): Science and Design}, Sci. China Phys. Mech. Astron. 63 (2020)
  219501.
\newblock \href {http://arxiv.org/abs/1810.09994} {\path{arXiv:1810.09994}},
  \href {https://doi.org/10.1007/s11433-018-9385-7}
  {\path{doi:10.1007/s11433-018-9385-7}}.

\bibitem{GRAND:2025rps}
K.~Kotera, et~al., {The Hybrid Elevated Radio Observatory for Neutrinos (HERON)
  Project}, PoS ICRC2025 (2025) 1078.
\newblock \href {http://arxiv.org/abs/2507.04382} {\path{arXiv:2507.04382}},
  \href {https://doi.org/10.22323/1.501.1078} {\path{doi:10.22323/1.501.1078}}.

\bibitem{Lazar:2023gem}
J.~Lazar, et~al., {Prometheus: An Open-Source Neutrino Telescope Simulation},
  Comput. Phys. Commun. 302 (2024) 109247.
\newblock \href {http://arxiv.org/abs/2304.14526} {\path{arXiv:2304.14526}},
  \href {https://doi.org/10.1016/j.cpc.2024.109247}
  {\path{doi:10.1016/j.cpc.2024.109247}}.

\bibitem{vanSanten:2022wss}
J.~van Santen, B.~A. Clark, R.~Halliday, S.~Hallmann, A.~Nelles, {toise: a
  framework to describe the performance of high-energy neutrino detectors},
  JINST 17~(08) (2022) T08009.
\newblock \href {http://arxiv.org/abs/2202.11120} {\path{arXiv:2202.11120}},
  \href {https://doi.org/10.1088/1748-0221/17/08/T08009}
  {\path{doi:10.1088/1748-0221/17/08/T08009}}.

\bibitem{IceCube:2021xar}
R.~Abbasi, et~al., {IceCube Data for Neutrino Point-Source Searches: Years
  2008--2018}, Astrophys. J. Lett. 923 (2021) L3.
\newblock \href {http://arxiv.org/abs/2101.09836} {\path{arXiv:2101.09836}},
  \href {https://doi.org/10.3847/2041-8213/ac2c7b}
  {\path{doi:10.3847/2041-8213/ac2c7b}}.

\bibitem{Koehne:2013gpa}
J.~H. Koehne, K.~Frantzen, M.~Schmitz, T.~Fuchs, W.~Rhode, D.~Chirkin,
  J.~Becker~Tjus, {PROPOSAL: A tool for propagation of charged leptons},
  Comput. Phys. Commun. 184 (2013) 2070--2090.
\newblock \href {https://doi.org/10.1016/j.cpc.2013.04.001}
  {\path{doi:10.1016/j.cpc.2013.04.001}}.

\bibitem{Dunsch:2018nsc}
M.~Dunsch, J.~Soedingrekso, A.~Sandrock, M.~Meier, T.~Menne, W.~Rhode, {Recent
  Improvements for the Lepton Propagator PROPOSAL}, Comput. Phys. Commun. 242
  (2019) 132--144.
\newblock \href {http://arxiv.org/abs/1809.07740} {\path{arXiv:1809.07740}},
  \href {https://doi.org/10.1016/j.cpc.2019.03.021}
  {\path{doi:10.1016/j.cpc.2019.03.021}}.

\bibitem{Groom:2001kq}
D.~E. Groom, N.~V. Mokhov, S.~I. Striganov, {Muon stopping power and range
  tables 10-MeV to 100-TeV}, Atom. Data Nucl. Data Tabl. 78 (2001) 183--356.
\newblock \href {https://doi.org/10.1006/adnd.2001.0861}
  {\path{doi:10.1006/adnd.2001.0861}}.

\bibitem{nuflux}
{IceCube Collaboration}, \href{https://github.com/icecube/nuflux}{{nuflux:
  Atmospheric neutrino flux library}} (2022).
\newline\urlprefix\url{https://github.com/icecube/nuflux}

\bibitem{Fedynitch:2015zma}
A.~Fedynitch, R.~Engel, T.~K. Gaisser, F.~Riehn, T.~Stanev, {Calculation of
  conventional and prompt lepton fluxes at very high energy}, EPJ Web Conf. 99
  (2015) 08001.
\newblock \href {http://arxiv.org/abs/1503.00544} {\path{arXiv:1503.00544}},
  \href {https://doi.org/10.1051/epjconf/20159908001}
  {\path{doi:10.1051/epjconf/20159908001}}.

\bibitem{Fedynitch:2018cbl}
A.~Fedynitch, F.~Riehn, R.~Engel, T.~K. Gaisser, T.~Stanev, {Hadronic
  interaction model sibyll 2.3c and inclusive lepton fluxes}, Phys. Rev. D
  100~(10) (2019) 103018.
\newblock \href {http://arxiv.org/abs/1806.04140} {\path{arXiv:1806.04140}},
  \href {https://doi.org/10.1103/PhysRevD.100.103018}
  {\path{doi:10.1103/PhysRevD.100.103018}}.

\bibitem{Dembinski:2017zsh}
H.~P. Dembinski, R.~Engel, A.~Fedynitch, T.~Gaisser, F.~Riehn, T.~Stanev,
  {Data-driven model of the cosmic-ray flux and mass composition from 10 GeV to
  $10^{11}$ GeV}, PoS ICRC2017 (2018) 533.
\newblock \href {http://arxiv.org/abs/1711.11432} {\path{arXiv:1711.11432}},
  \href {https://doi.org/10.22323/1.301.0533} {\path{doi:10.22323/1.301.0533}}.

\bibitem{Riehn:2019jet}
F.~Riehn, R.~Engel, A.~Fedynitch, T.~K. Gaisser, T.~Stanev, {Hadronic
  interaction model Sibyll 2.3d and extensive air showers}, Phys. Rev. D
  102~(6) (2020) 063002.
\newblock \href {http://arxiv.org/abs/1912.03300} {\path{arXiv:1912.03300}},
  \href {https://doi.org/10.1103/PhysRevD.102.063002}
  {\path{doi:10.1103/PhysRevD.102.063002}}.

\bibitem{Yanez:2023lsy}
J.~P. Ya\~nez, A.~Fedynitch, {Data-driven muon-calibrated neutrino flux}, Phys.
  Rev. D 107~(12) (2023) 123037.
\newblock \href {http://arxiv.org/abs/2303.00022} {\path{arXiv:2303.00022}},
  \href {https://doi.org/10.1103/PhysRevD.107.123037}
  {\path{doi:10.1103/PhysRevD.107.123037}}.

\bibitem{Braun:2008bg}
J.~Braun, J.~Dumm, F.~De~Palma, C.~Finley, A.~Karle, T.~Montaruli, {Methods for
  point source analysis in high energy neutrino telescopes}, Astropart. Phys.
  29 (2008) 299--305.
\newblock \href {http://arxiv.org/abs/0801.1604} {\path{arXiv:0801.1604}},
  \href {https://doi.org/10.1016/j.astropartphys.2008.02.007}
  {\path{doi:10.1016/j.astropartphys.2008.02.007}}.

\bibitem{IceCube:2019cia}
M.~G. Aartsen, et~al., {Time-Integrated Neutrino Source Searches with 10 Years
  of IceCube Data}, Phys. Rev. Lett. 124~(5) (2020) 051103.
\newblock \href {http://arxiv.org/abs/1910.08488} {\path{arXiv:1910.08488}},
  \href {https://doi.org/10.1103/PhysRevLett.124.051103}
  {\path{doi:10.1103/PhysRevLett.124.051103}}.

\bibitem{Astropy:2013muo}
T.~P. Robitaille, et~al., {Astropy: A community Python package for astronomy},
  Astron. Astrophys. 558 (2013) A33.
\newblock \href {http://arxiv.org/abs/1307.6212} {\path{arXiv:1307.6212}},
  \href {https://doi.org/10.1051/0004-6361/201322068}
  {\path{doi:10.1051/0004-6361/201322068}}.

\bibitem{Astropy:2018gua}
A.~M. Price-Whelan, et~al., {The Astropy Project: Building an Open-science
  Project and Status of the v2.0 Core Package}, Astron. J. 156 (2018) 123.
\newblock \href {http://arxiv.org/abs/1801.02634} {\path{arXiv:1801.02634}},
  \href {https://doi.org/10.3847/1538-3881/aabc4f}
  {\path{doi:10.3847/1538-3881/aabc4f}}.

\bibitem{pint}
H.~E. Grecco, et~al., \href{https://pint.readthedocs.io}{{pint: Physical
  quantities in Python}} (2023).
\newline\urlprefix\url{https://pint.readthedocs.io}

\bibitem{h5py}
A.~Collette, et~al., \href{https://www.h5py.org}{{h5py: A Pythonic interface to
  the HDF5 binary data format}} (2023).
\newline\urlprefix\url{https://www.h5py.org}

\bibitem{Virtanen:2019joe}
P.~Virtanen, et~al., {SciPy 1.0: Fundamental Algorithms for Scientific
  Computing in Python}, Nature Meth. 17 (2020) 261--272.
\newblock \href {http://arxiv.org/abs/1907.10121} {\path{arXiv:1907.10121}},
  \href {https://doi.org/10.1038/s41592-019-0686-2}
  {\path{doi:10.1038/s41592-019-0686-2}}.

\bibitem{Harris:2020xlr}
C.~R. Harris, et~al., {Array programming with NumPy}, Nature 585 (2020)
  357--362.
\newblock \href {http://arxiv.org/abs/2006.10256} {\path{arXiv:2006.10256}},
  \href {https://doi.org/10.1038/s41586-020-2649-2}
  {\path{doi:10.1038/s41586-020-2649-2}}.

\end{thebibliography}

\pagebreak

\appendix

\section{Software Architecture}
\label{app:architecture}

\spore{} is organized into five top-level sub-packages.

\paragraph{\texttt{spore.conventions}}
Defines the coordinate and unit primitives used throughout the package.
\texttt{SkyCoordinate} (declination, right ascension) and \texttt{LocalCoordinate} (zenith, azimuth) are lightweight dataclasses that validate their arguments on construction: out-of-range declinations and zenith angles raise, while right ascensions and azimuths outside $[0, 2\pi)$ are wrapped with a warning.
Coordinate transforms between the equatorial and local frames are implemented using \texttt{astropy.coordinates} \citep{Astropy:2013muo,Astropy:2018gua}.
A single shared \texttt{pint.UnitRegistry} instance~\citep{pint} is exported and imported by all modules, ensuring that unit comparisons and arithmetic are globally consistent.
Energies are stored internally in gigaelectronvolts (GeV); effective areas in cm$^2$; angular errors in radians; and livetimes in seconds.

\paragraph{\texttt{spore.physics}}
Defines the neutrino species enumeration and event morphology registry.
The \texttt{Neutrino} enum lists all six standard species ($\nu_e, \bar\nu_e, \nu_\mu, \bar\nu_\mu, \nu_\tau, \bar\nu_\tau$) together with their PDG Monte Carlo particle identifiers.
The module-level list \texttt{neutrinos} provides the canonical iteration order used by all samplers when summing over flavors.
The \texttt{Morphology} class maintains a registry of known event morphology labels; standard labels (\texttt{track}, \texttt{cascade}) are pre-registered, and additional morphologies (e.g.\ double-cascade topologies) can be added via \texttt{Morphology.register} before loading a detector response.
The \texttt{DetectorResponse} loader iterates the registry when scanning an HDF5 file and warns about any groups that match the morphology naming convention but are not registered.

\paragraph{\texttt{spore.source}}
Implements flux models and source geometries.
\texttt{Flux} stores a per-species normalization and a \texttt{Distribution} subclass; built-in distributions cover energy-only (\texttt{PowerLaw}, \texttt{TabulatedEnergyFlux}), energy and declination (\texttt{TabulatedEnergyDecFlux}), and energy, declination, and right ascension (\texttt{TabulatedEnergyDecRAFlux}) spectral shapes.
The \texttt{uses\_ra} property of \texttt{ExtendedSource} is inferred automatically from the flux: if any species uses a \texttt{TabulatedEnergyDecRAFlux} distribution the sampler builds a full $(\delta, \alpha, E)$ grid without any additional configuration.
TOML-based configuration is supported via \texttt{Flux.from\_config} and \texttt{PointSource.from\_config} (see \cref{app:toml}).

\paragraph{\texttt{spore.detector}}
Manages the loading and interpolation of instrument response functions.
The top-level \texttt{Detector} class bundles a \texttt{DetectorResponse} with an \texttt{EarthCoordinate} and a reference epoch.
\texttt{DetectorResponse} exposes three callable dictionaries keyed by registered morphology name: the effective area $A_\mathrm{eff}(\zeta, E)$ in cm$^2$; the PSF inverse CDF $\psi(E, u)$ (or \texttt{None} if absent from the file); and the energy resolution inverse CDF $\Delta(u)$ (or \texttt{None} if absent).
IRFs are loaded from HDF5 files (\cref{app:hdf5}) via \texttt{h5py}~\citep{h5py} and interpolated with \texttt{scipy.interpolate}~\citep{Virtanen:2019joe}; all grids and sampling tables are \texttt{numpy} arrays~\citep{Harris:2020xlr}.

\paragraph{\texttt{spore.event\_sampling}: samplers}
Contains the sampler classes, the \texttt{Event} class, and supporting utilities.
The \texttt{SourceSampler} factory function accepts a detector (or list of detectors) and a source object and returns the appropriate sampler: \texttt{PointSourceEventSampler} for a \texttt{PointSource} and \texttt{ExtendedSourceEventSampler} for an \texttt{ExtendedSource}.
Both sampler classes share the abstract base class \texttt{EventSampler}, which implements the public \texttt{sample\_events} and \texttt{expected\_events} entry points---including the broadcasting of per-detector arguments---and requires each subclass to supply the corresponding single-detector methods.

\paragraph{\texttt{spore.event\_sampling}: smearing}
The shared utility function \texttt{smear\_truth} applies PSF and energy smearing to a true (direction, energy) pair, returning a reconstructed direction, reconstructed energy, and per-event angular uncertainty.
Where the response uses a joint smearing IRF, the band is selected by local zenith, so calling this utility directly requires either an explicit zenith or an epoch from which to compute one; the sampler classes supply it automatically, and it is well defined from the declination alone only for a detector at a geographic pole.
The (inverse-CDF spline, normalization) pair computed at construction time is cached per (epoch, morphology) key, so repeated calls for pseudo-experiments at the same epoch incur negligible overhead.

\paragraph{\texttt{spore.event\_sampling}: multiple detectors and livetime}
Both sampler classes accept a \texttt{List[Detector]} to enable multi-detector joint analyses: each detector's events are sampled independently and tagged with an integer detector index before being returned in a single combined list.
The \texttt{GoodRunList} class encodes a set of valid-data intervals loaded from IceCube-style uptime CSV files; it can be constructed from a single file, a directory of files, or a list of either.
All per-call arguments (\texttt{deltat}, \texttt{t}, \texttt{delta\_clip}, and \texttt{grl}) may be passed as scalars broadcast to every detector, or as lists of per-detector values with length matching the detector list.

\section{HDF5 Response File Format}
\label{app:hdf5}

A \spore{}-compatible HDF5 file uses a hierarchical format in which each supported morphology occupies a top-level group named after the morphology label (e.g.\ \texttt{track}, \texttt{cascade}).
Each morphology group contains one required subgroup and up to three optional subgroups; \cref{tab:hdf5} summarizes the schema.
Morphology groups absent from the file are simply not loaded.
If the angular response or energy resolution subgroup is absent, the corresponding reconstructed quantities are set to \texttt{NaN} in the sampled events, allowing the package to be used for rate-only studies where only the effective area is available.

\begin{table*}[t]
\centering
\caption{HDF5 instrument response file schema.
  Each subgroup lives inside a top-level morphology group (e.g.\ \texttt{track/} or \texttt{cascade/}).
  Optional subgroups are indicated by $\dagger$; their absence causes the corresponding reconstructed field to be filled with \texttt{NaN}.}
\label{tab:hdf5}
\small
\begin{tabular}{lll}
\toprule
Subgroup & Axes & Units \\
\midrule
\texttt{effective\_area} & \texttt{zeniths}, \texttt{energies} & cm$^2$ \\
\texttt{angular\_response}$^\dagger$ & \texttt{energies}, \texttt{quantile} & radians \\
\texttt{energy\_resolution}$^\dagger$ & \texttt{quantile} & dimensionless \\
\texttt{smearing}$^\dagger$ & \texttt{log10e\_true}, \texttt{dec}, \texttt{log10e\_reco}, \texttt{psf}, \texttt{ang\_err} & fractional counts \\
\bottomrule
\end{tabular}
\end{table*}

\paragraph{Effective area (\texttt{effective\_area}): storage}
A two-dimensional array of shape $(n_E, n_\zeta)$ storing $A_\mathrm{eff}(\zeta, E)$ in cm$^2$, or of shape $(6, n_E, n_\zeta)$ for responses that resolve the six neutrino species separately (required where flavors have significantly different detection efficiencies, as for HESE cascades at high energy).
The axis datasets \texttt{zeniths} (length $n_\zeta$, in radians) and \texttt{energies} (length $n_E$, in GeV) give the grid-centre values.

\paragraph{Effective area: pre-processing}
Two pre-processing steps are applied at load time and are worth stating explicitly, since both modify the tabulated values before any sampling takes place.
First, for each zenith column all energy bins outside the longest contiguous run of non-zero bins are zeroed (\texttt{trim\_isolated}, default on), which removes isolated low-statistics bins at the edges of the sensitivity range.
Second, each zenith column is Gaussian-smoothed in $(\ln E,\, \ln A_\mathrm{eff})$ space with a kernel of width \texttt{smoothing\_sigma} energy bins (default $1.5$), which suppresses bin-to-bin Monte Carlo noise in the high-energy tail.
The two steps are independent, and either may be overridden per call or per file through a \texttt{meta} group in the HDF5 file; setting \texttt{smoothing\_sigma} to zero disables smoothing while leaving trimming in place.
The appropriate smoothing width is IRF-dependent, and users should plot the loaded effective area and confirm it is faithful to the tabulated input before using it for analysis.

\paragraph{Effective area: interpolation}
$A_\mathrm{eff}(\zeta, E)$ is then evaluated by PCHIP interpolation along $\ln E$ within each $\cos\zeta$ column, followed by linear interpolation in $\cos\zeta$ between the two bracketing columns, with both stages carried out on $\ln A_\mathrm{eff}$.
The $\cos\zeta$ grid is linearly extrapolated in this log space to the hard boundaries $\cos\zeta = \pm 1$ so that queries at the poles return finite values rather than zero.
Queries outside the tabulated energy range, and values falling below one percent of the smallest non-zero tabulated entry, return zero.

\paragraph{Angular response (\texttt{angular\_response}, optional)}
A two-dimensional array of shape $(n_E, n_u)$ storing the inverse CDF of the PSF deflection angle as a function of energy.
At energy bin $k$, the value at quantile $u_l$ is the angle $\psi$ (in radians) such that a fraction $u_l$ of events are reconstructed within $\psi$ of the true direction.
The axis datasets are \texttt{energies} (length $n_E$, in GeV) and \texttt{us} (length $n_u$, uniformly spaced in $[0,1]$).
This representation enables direct inverse-CDF sampling without rejection: drawing $u \sim \mathrm{Uniform}(0,1)$ and evaluating $\psi(E, u)$ produces a deflection angle distributed as the true PSF.
If this subgroup is absent, reconstructed directions and angular uncertainties are set to \texttt{NaN}.

\paragraph{Energy resolution (\texttt{energy\_resolution}, optional)}
A one-dimensional array of length $n_u$ storing the inverse CDF of $\Delta \equiv \ln(E_\mathrm{reco}/E_\mathrm{true})$.
For tracks this captures the stochastic loss fluctuations along the muon track; for cascades it captures the shower energy measurement uncertainty.
The axis dataset \texttt{us} gives the corresponding quantile values in $[0,1]$.
A reconstructed energy is obtained as $E_\mathrm{reco} = E_\mathrm{true} \cdot \exp[\mathrm{spl}(u)]$, where $\mathrm{spl}$ is a PCHIP spline fitted to the stored inverse CDF and $u \sim \mathrm{Uniform}(0,1)$.
If this subgroup is absent, reconstructed energies are set to \texttt{NaN}.

\paragraph{Joint smearing (\texttt{smearing}, optional)}
A five-dimensional array of fractional counts encoding $P(E_\mathrm{reco}, \psi, \sigma_\psi \mid E_\mathrm{true}, \delta)$, in the binned form released with the IceCube ten-year point-source sample~\citep{IceCube:2021xar}.
When present, this subgroup supersedes \texttt{angular\_response} and \texttt{energy\_resolution} for that morphology: a single inverse-CDF draw over the flattened joint histogram returns all three reconstructed quantities at once, and the \texttt{delta\_clip} argument has no effect.
The declination axis is converted to a local zenith axis at load time using the identity $\zeta = \arccos(-\sin\delta)$ appropriate to a South Polar detector, so a response file written in this format is tied to the detector latitude of the release it came from.
Note that a file supplying \texttt{smearing} but no \texttt{energy\_resolution} still triggers the loader warning that reconstructed energies will be \texttt{NaN}; the warning is spurious in that case, as the joint sampler supplies them.

Builder scripts in the \texttt{scripts/} directory of the repository demonstrate how to construct compliant HDF5 files from the IceCube ten-year \citep{IceCube:2021xar} and HESE 7.5-year \citep{IceCube:2020wum} public Monte Carlo releases.

\section{TOML Configuration Reference}
\label{app:toml}

A \texttt{PointSource} can be fully specified by a TOML file with two required tables: \texttt{[location]} and \texttt{[flux]}.
Every \texttt{from\_config} constructor accepts either a mapping or a path to such a file, which is parsed with \texttt{tomllib} from the standard library; a \texttt{Flux} may be built from a bare flux file or from the \texttt{[flux]} table of a source file.

\paragraph{\texttt{[location]}}
\begin{description}
  \item[\texttt{right\_ascension}] Right ascension in decimal degrees (J2000).
  \item[\texttt{declination}] Declination in decimal degrees (J2000).
\end{description}

\paragraph{\texttt{[flux]}}
\begin{description}
  \item[\texttt{norm\_per\_species}] Per-species normalization in
    GeV$^{-1}$\,cm$^{-2}$\,s$^{-1}$ at the pivot energy.
    For a total $(\nu+\bar\nu)$ track flux of $\Phi_0$ at the pivot, set
    \texttt{norm\_per\_species = 0.5 * Phi\_0}.
  \item[\texttt{gamma}] Spectral index $\gamma$ of the power law
    $\Phi \propto (E/E_\mathrm{pivot})^{-\gamma}$.
  \item[\texttt{pivot}] Pivot energy in GeV.
  \item[\texttt{emin}] Lower energy bound in GeV (default $10^2$~GeV).
  \item[\texttt{emax}] Upper energy bound in GeV (default $10^6$~GeV).
\end{description}

An example configuration file:

\begin{verbatim}
[location]
right_ascension = 50.0   # degrees, J2000
declination     =  5.0   # degrees, J2000

[flux]
norm_per_species = 1e-18  # GeV^-1 cm^-2 s^-1 per species
gamma      = 2.0
pivot      = 1e5          # GeV
emin       = 1e2          # GeV
emax       = 1e6          # GeV
\end{verbatim}

Detector configuration can likewise be specified as a TOML file with \texttt{[properties]} and \texttt{[response]} tables, or as the equivalent Python dictionary, passed to \texttt{Detector.from\_config}.

\end{document}